\documentclass[journal=jctcce,manuscript=article]{achemso}
\setkeys{acs}{articletitle = true}

\usepackage{chemformula} 
\usepackage[T1]{fontenc} 
\usepackage{amsmath}
\usepackage{mathtools}
\usepackage{relsize}

\newcommand{\eu}{\mathrm{e}^}

\newcommand{\ders}[3][]{\frac{\rmd^2{#2}}{\rmd{#3}^2}}
\newcommand{\pder}[3][]{\frac{\partial^{#1}{#2}}{\partial{#3}^{#1}}}
\newcommand{\pders}[3]{\frac{\partial^2{#1}}{\partial{#2}\partial{#3}}}
\newcommand{\red}{\color{black}{}}
\newcommand{\orange}{\color{black}{}}

\newcommand{\Ref}[1]{ref~\citenum{#1}}

\newcommand{\rmd}{\mathrm{d}}
\newcommand{\thalf}{{\ensuremath{\tfrac{1}{2}}}} 

\author{Pierre Winter}
\author{Jeremy O. Richardson}
\affiliation{Laboratory of Physical Chemistry, ETH Z{\"u}rich, 8093 Z{\"u}rich, Switzerland}
\email{jeremy.richardson@phys.chem.ethz.ch}

\title{Divide-and-Conquer Method for Instanton Rate Theory}

\abbreviations{DaC}

\begin{document}

\begin{abstract}
\noindent
Ring-polymer instanton theory has been developed to simulate the quantum dynamics of molecular systems at low temperatures.
Chemical reaction rates can be obtained by locating the dominant tunneling pathway and analyzing fluctuations around it.
In the standard method,
calculating the fluctuation terms involves the diagonalization of a large matrix, which can be unfeasible for large systems with a high number of ring-polymer beads.
Here we present a method for computing the instanton fluctuations 
with a large reduction in computational scaling.
This method is applied to three reactions described by fitted, analytic and on-the-fly \orange{ab initio} potential-energy surfaces
and is shown to be numerically stable for the calculation of thermal reaction rates even at very low temperature.
\end{abstract}


\section{Introduction}

Quantum tunneling is an important dynamical phenomenon
which must be taken into account in order to accurately model certain chemical reactions, especially at low temperature. \cite{BellBook}
Instanton theory has been developed to include such effects by using a semiclassical approximation to the exact rate constant \cite{Miller1975semiclassical,AdiabaticGreens,InstReview}.
The instanton uniquely defines the dominant tunneling pathway through a potential-energy barrier and can be used to directly calculate reaction rates.
This theory can be applied to molecular systems in full dimensionality
using computational methods based on
discretization of the tunneling pathway into ring-polymer beads,
giving rise to ring-polymer instanton (RPI) theory \cite{Andersson2009Hmethane,InstReview, RPInst, Rommel2011locating, Perspective}.
This theory enables efficient calculations of reaction rates in the deep tunneling regime, where classical transition-state theory (TST) or shallow tunneling theories are not valid \cite{BellBook, Wigner1932parabolic,Haenggi1990rate}.
RPI theory has been applied successfully to gas-phase chemistry \cite{Asgeirsson2018instanton, Kaestner2014review, HCH4}, molecular cluster rearrangements \cite{tunnel, *water, hexamerprism, formic, i-wat2}, surface reactions \cite{Jonsson2011surface, Lamberts2017ice, *Meisner2017water, Senevirathne2017ice}, and hydrogen transfer in enzymes \cite{Rommel2012enzyme}.

RPI theory takes as input the potential-energy surface (PES) as well as its first and second derivatives
at points along the tunneling trajectory. 
Recent advances in optimization techniques have allowed for speed-ups in locating the instanton,
requiring fewer calculations of the PES \cite{Asgeirsson2018instanton}.
Further improvements in efficiency have been made with the introduction of machine-learning methodology
which can be used to create a local PES around the instanton using only a small number of \textit{ab initio} calculations \cite{GPR,Cooper2018interpolation}.
The application of these techniques allows for RPI theory to be used with high-level electronic-structure methods,
with a computational cost which in the best cases is not too much higher than TST. 

Due to these advances, the computational bottleneck for the instanton method may sometimes be the diagonalization of the ring-polymer Hessian when calculating fluctuations around the instanton.
Especially when using a large number of beads, which is necessary at low temperatures,
diagonalization of this matrix requires both powerful processing power and a large amount of memory.
In this paper we briefly review instanton rate theory from a computational perspective,
introduce a novel method for calculating the instanton fluctuations, and finally apply this method to three different reaction systems.
We will see that this new approach gives the same results as current methods, but with a large reduction in computation time,
which we show scales linearly with the number of beads.

\section{Instanton Rate Theory}

Instanton theory is based on a semiclassical approximation \cite{InstReview} to the quantum reaction rate \cite{Miller1983rate}
written in terms of the path-integral representation of quantum mechanics \cite{Feynman}.
The instanton is a periodic orbit \cite{Miller1975semiclassical} of imaginary time \red{$\tau = \beta \hbar$} and exists only below the crossover temperature, 
determined by the magnitude of the imaginary barrier frequency, $\omega_\text{b}$, as $T_\mathrm{c}=\hbar\omega_\text{b}/2\pi k_\text{B}$. 
Above $T_\mathrm{c}$, the instanton collapses at the barrier top
and the rate \red{constant} can be calculated by other methods \cite{BellBook, Wigner1932parabolic}.

Within the semiclassical approximation, the imaginary-time kernel
for a system of $f$ nuclear degrees of freedom is given by \cite{GutzwillerBook}
\begin{equation}
K_{SC}(x',x'',\tau) \sim \sum_\mathrm{traj.} \sqrt\frac{C}{(2\pi\hbar)^f} \, \eu{-S/\hbar}
\label{eqn:kernel_full}
\end{equation}
Rather than taking an integral over all possible paths, which would give the exact result, \cite{Feynman}
this approximation allows us to treat a sum of classical trajectories which extremize the action
\begin{equation}
S \equiv S(x', x'', \tau) = \int_0^{\tau} \left[\thalf m \Vert\dot{x}\Vert^2 + V(x)\right] \rmd t
\label{eqn:action}
\end{equation}
The trajectory describes the dynamics of a particle moving from an initial point $x'$ to a final point $x''$ in imaginary time $\tau$ on the potential-energy surface $V(x)$.
Note that $x$ is an $f$-dimensional vector and we use mass weighting such that all degrees of freedom have the same mass, $m$.
The van-Vleck \cite{vanVleck1928correspondence} prefactor, $C$, is the determinant of the \red{second derivative 
of the action}
with respect to the initial and final endpoints of a given trajectory \cite{GutzwillerBook}
\begin{equation}
C = \left| - \pders{S}{x'}{x''} \right|
\label{eqn:C_mat}
\end{equation}

In RPI theory, the instanton trajectory with imaginary time $\tau$ is discretized by a ring polymer with $N$ beads.
Although not generally necessary, \cite{Rommel2011grids,GoldenRPI}
here we assume even spacing of the imaginary time between each bead.
It is now well established
that the 
discretized instanton can be located with a saddle point search on the ring-polymer potential \cite{Andersson2009Hmethane, RPInst, Rommel2011locating}.

An expression for the thermal rate \red{constant} can be derived 
by replacing the kernels appearing in the exact rate expression \cite{Miller1983rate}
by their semiclassical approximation, eq \ref{eqn:kernel_full},
and performing a number of steepest-descent integrals \cite{InstReview}.
The thermal rate \red{constant} at temperature $\beta = 1 / k_\text{B} T$ for a reaction with symmetry number $\sigma$ is given by the ring-polymer instanton expression
\begin{equation}
k_{\mathrm{RPI}} = \frac{\sigma}{2 \pi \beta \hbar} \frac{Z^{\mathrm{tr}}_\mathrm{RPI} Z^{\mathrm{rot}}_\mathrm{RPI} Z^{\mathrm{vib}}_\mathrm{RPI}}
{Z^{\mathrm{tr}}_\mathrm{r} Z^{\mathrm{rot}}_\mathrm{r} Z^{\mathrm{vib}}_\mathrm{r}} \, \eu{-S/\hbar}
\label{eqn:kt}
\end{equation}
This is similar to the formula for classical Eyring TST, \cite{Eyring1938rate}
except that the exponent depends on the action of the instanton rather than the barrier height.
Note also that the partition functions in the numerator refer to the ring-polymer representation of the instanton rather than the classical transition state and the partition functions in the denominator refer to the collapsed ring-polymer representations of the reactants.
\red{Separability of rotational and vibrational modes is also assumed in this formulation.}
The ring-polymer instanton approach gives thermal reaction rates which are in good agreement with other high-accuracy methods \cite{Andersson2009Hmethane,MUSTreview,HCH4,DMuH}. The vibrational partition function of the instanton can be calculated in a variety of ways and it is the purpose of this paper to discover the most efficient and reliable approach for this task.

\subsection{Instanton Vibrational Partition Function} \label{Zvib}

There are two original formulations of instanton theory, \cite{Uses_of_Instantons,Miller1975semiclassical}
which have been shown to be identical \cite{Althorpe2011ImF}.
Recently, instanton rate theory has been re-derived from first principles \cite{AdiabaticGreens}
which leads to yet more alternative expressions \cite{InstReview}.
These approaches are all formally equivalent, although they differ in how they numerically calculate the instanton vibrational partition function in eq \ref{eqn:kt}.
There are three general methods for calculating $Z^{\mathrm{vib}}_\mathrm{RPI}$,
which we describe in the following.

\subsubsection{Ring-Polymer Eigenvalues} \label{RPE}

The standard approach for calculating the instanton vibrational partition function is to make use of a normal-mode analysis of the
$N$-bead ring-polymer
discretization of the instanton \red{and to take steepest-descent integrals over all beads simultaneously, which gives} 
\cite{RPInst,Andersson2009Hmethane,Rommel2011locating}
\begin{equation}
Z^{\mathrm{vib}}_\mathrm{RPI} = N^{f_0+1} \left( \frac{2 \pi m B_N}{\beta_N \hbar^2} \right)^{1/2} \sideset{}{'}\prod_k \left| \frac{1}{\beta_N \hbar \eta_k} \right|
\label{eqn:Z_RPevals}
\end{equation}
Here $f_0$ is the combined number of translational and rotational degrees of freedom, $\beta_N=\beta/N$, and $\eta_k^2$ are the eigenvalues of the mass-weighted ring-polymer Hessian. $B_N$ is a measure of the squared displacement
between neighboring ring-polymer beads, as defined previously \cite{InstReview}.

Because the instanton sits on a first-order saddle-point of the ring-polymer potential,
the corresponding Hessian will have one negative eigenvalue and for this reason the product in eq \ref{eqn:Z_RPevals} is performed over absolute values of $\eta_k$.
The ring-polymer Hessian also produces one zero eigenvalue corresponding to a cyclic permutation of the beads and
$f_0$ additional zero eigenvalues corresponding to translations and rotations in the system.
The prime indicates omission of all these zero eigenvalues from the product.

Because this is the most commonly applied approach and has been thoroughly tested,
the ring-polymer eigenvalues (RPE) method will be used as a benchmark with which to compare other methods in this paper.
The drawback with the RPE method is the large computational cost needed to diagonalize a large $Nf \times Nf$ matrix which scales roughly as $\mathcal{O}(N^3f^3)$ \cite{IntelMKL, KaufmanScaling, LAPACK}.
Since the ring-polymer Hessian is banded except for off-diagonal elements connecting the first and last beads, this scaling could in principle be improved by using \red{algorithms that exploit the
sparsity of such matrices}. However, in subsections \ref{stabil} and \ref{PRP} we will present methods which scale even better than sparse matrix \red{algorithms}.
In this paper, we will not be concerned with scaling with respect to $f$ since this is inherent to the system under study
and certainly cannot be lower than the $\mathcal{O}(f^3)$ of a TST calculation.

\subsubsection{Stability Parameters} \label{stabil}

It is also possible to calculate $Z^{\mathrm{vib}}_\mathrm{RPI}$ using stability parameters (SP) \cite{Miller1975semiclassical,Chapman1975rates}.
The vibrational partition function in this case takes the form
\begin{equation}
Z^{\mathrm{vib}}_\mathrm{RPI} =  \beta \sqrt{2 \pi\hbar} \left( - \frac{\mathrm{d}E}{\mathrm{d}\tau}\right) ^{1/2} \sideset{}{'}\prod_j  \frac{1}{2 \sinh(u_j /2)}
\label{eqn:Z_stabil}
\end{equation}
where $u_j$ are the stability parameters \cite{GutzwillerBook} of the instanton trajectory and $E$ defines the energy at which the imaginary-time instanton period, $\tau$, equals $\beta \hbar$.
The prime indicates that the mode corresponding to the reaction coordinate and those corresponding to the translations and rotations in the system are omitted from the product since their stability parameters are zero.


The \red{total} derivative $\frac{\mathrm{d}E}{\mathrm{d}\tau}$ can be computed with finite differences
by comparing the energy of two or three instantons optimized at slightly different temperatures. \cite{Kryvohuz2011rate}
This approach is however found to be numerically unstable at very low temperatures \orange{and requires the optimization of many more instantons}. 

It can also be defined
as $\frac{\mathrm{d}E}{\mathrm{d}\tau} = - \left( \ders{W}{E} \right)^{-1}$,
in terms of the Legendre transform of the action, $W(x', x'', E) = S(x', x'', \tau) - E \tau$, \orange{along the trajectory.}
As in eq \ref{eqn:action}, the action is treated as a function of the endpoints of the trajectory, $x'$ and $x''$, and imaginary time $\tau$,
\orange{and the expression is evaluated for the closed trajectory $x'=x''$.}
The expression for $\ders{W}{E}$ is given by the chain rule in eq 37 of ref \citenum{InstReview} and can be further substituted with equations A5-A7 of the same reference. Rearranging terms gives
\begin{equation}
\frac{\mathrm{d}E}{\mathrm{d}\tau} =  \pder[2]{S}{\tau}
\left[ 1 +
\pders{S}{\tau}{x} 
\left(  \pder[2]{S}{\tau} \pders{S}{x}{x} - \pders{S}{x}{\tau} \pders{S}{\tau}{x} \right)^{-1}  \pders{S}{x}{\tau} \right ] ^{-1}
\label{eqn:d2WdE2}
\end{equation}
where we have used the notation $x' = x'' = x$ such that $\pders{S}{x}{x} = \pders{S}{x'}{x'} + \pders{S}{x'}{x''} + \pders{S}{x''}{x'} + \pders{S}{x''}{x''}$ and $\pders{S}{x}{\tau} = \pders{S}{x'}{\tau} + \pders{S}{x''}{\tau}$.
The total derivative is thus defined according to the chain rule, effectively allowing the endpoints \red{of the trajectory} to change with $\tau$.

\orange{A third method for obtaining $\frac{\mathrm{d}E}{\mathrm{d}\tau}$ is given by \citeauthor{KastnerAction} in ref \citenum{KastnerAction}. 
Their approach requires calculating the total derivative $\frac{\mathrm{d}x_{i}}{\mathrm{d}\tau}$, which can be found by solving a system of linear equations for the closed instanton periodic orbit and where $x_i$ is an optimized bead position. 
They propose to choose the index $i$ to refer to the bead at the reactant turning point of the instanton in order to avoid numerical instability.

Disregarding the finite difference method, it is not clear which of the two methods for calculating $\frac{\mathrm{d}E}{\mathrm{d}\tau}$ is best. The approach used in this paper instead combines both ideas in a way which improves numerical stability. Here eq \ref{eqn:d2WdE2} will be used with $\pder[2]{S}{\tau}$ given by
\begin{equation}
\begin{aligned}
\pder[2]{S}{\tau} &= \frac{m N^2}{\tau^3} \left\lVert x_i - x_{i-1}\right\rVert^2 - \frac{m N^2}{\tau^2} \left( x_i - x_{i-1}\right) \cdot \left( \pder {x_i}{\tau} - \pder{x_{i-1}}{\tau} \right) \\
& + \frac{1}{2} \left( \pder{V}{x_i} \cdot \pder{x_{i}}{\tau} + \pder{V}{x_{i-1}} \cdot \pder{x_{i-1}}{\tau} \right)
\label{eqn:d2Sdt2}
\end{aligned}
\end{equation}
and with $x_i$ chosen as the reactant turning point of the instanton. The partial derivatives $\pders{S}{x}{x}$ and  $\pders{S}{x}{\tau}$ are given by formulas from ref \citenum{GoldenRPI}. For larger systems this combined approach will be more favorable because, unlike the respective total derivatives, the partial derivatives $\pder {x_i}{\tau} $ and $\pder{x_{i-1}}{\tau}$ can be obtained using a banded-matrix linear solver.}
It should be noted that for a unimolecular reaction in the low-temperature limit, the instanton trajectory moves only very slightly with a change in $\tau$ and the rate plateaus to a constant.
In this plateau region, $\frac{\mathrm{d}E}{\mathrm{d}\tau}$ tends to $\pder[2]{S}{\tau}$ and approaches zero quickly. 
In order to avoid numerical errors when using eq \ref{eqn:d2WdE2}, the approximation $\frac{\mathrm{d}E}{\mathrm{d}\tau}  \approx \pder[2]{S}{\tau}$ should be used at low temperatures.

Obtaining the stability parameters, $u_j$, requires constructing the symplectic monodromy matrix, $\mathbf{M}(\tau)$,
which can be defined as the solution to the following differential equation: 
\orange{
\begin{equation}
 \frac{\mathrm{d}}{\mathrm{d}\tau} \mathbf{M}(\tau)
 =  \begin{pmatrix} 
      \mathbf{0} & \mathbf{m}^{-1}  \\
      \nabla ^2 \mathbf{V}(x(\tau)) & \mathbf{0}
   \end{pmatrix}
\mathbf{M}(\tau)
\label{eqn:monodromy}
\end{equation}}
with the initial conditions $\mathbf{M}(0) = \mathbf{I}$\@.
Here $\mathbf{m}=m\mathbf{I}$ is the $f \times f$ identity matrix weighted by the system mass
and \orange{$\nabla ^2 \mathbf{V}(x(\tau))$} is the $f \times f$ Hessian along the trajectory at imaginary-time $t$.
The monodromy matrix itself describes how the final position and momentum of a system depends on changes in the initial position and momentum \cite{Miller1975semiclassical}.
The corresponding eigenvalues of this non-symmetric matrix are of the form $\eu{\pm u_j}$,
which defines the stability parameters to be used in eq \ref{eqn:Z_stabil}. 
Numerical propagation of \orange{$\mathbf{M}(\tau)$ from $\tau=0$ to $\tau=\beta \hbar$} using eq \ref{eqn:monodromy} can be performed with the well known Runge-Kutta 4th-order method (SP-RK4).
This method has been used previously for instanton calculations \cite{Chapman1975rates,Ceotto2012instanton,Kryvohuz2011rate,Kryvohuz2012abinitio,*Kryvohuz2012instanton,*Kryvohuz2014KIE, KastnerAction, McConnell2017microcanonical, McConnell2017instanton}
but never at very low temperatures.
\citeauthor{KastnerAction} have also proposed an alternative approach to building $\mathbf{M}(\tau)$ using the second \red{derivative} of the action (SP-Action) \cite{KastnerAction}.
At very low temperatures, however, both the SP-RK4 and SP-Action methods completely break down and they can not produce reliable rate constants as shown in section \ref{Results and Discussion}. We find that with the SP approaches, the problem lies not so much in building the monodromy matrix but rather in diagonalizing it because it has eigenvalues of widely different magnitudes.

Despite the many approaches for building the monodromy matrix, \cite{DiLiberto2016monodromy} so far no solutions have been suggested for dealing with problems in diagonalizing this non-symmetric matrix for large values of $\tau$, i.e.\ at low temperatures.
In this regime, the eigenvalue pairs of $\mathbf{M}(\tau)$ differ by many orders of magnitude and diagonalization routines are unable to produce stable values of $u_j$. 
The numerical problems introduced when calculating $\frac{\mathrm{d}E}{\mathrm{d}\tau}$ and $u_j$ are independent of the number of beads used to discretize the instanton trajectory and increasing $N$ does not avoid the breakdown of the SP method.
Use of the SP method to calculate the instanton vibrational partition function is therefore only
numerically stable for trajectories with small $\tau$, which correspond to reactions at higher temperatures.
It would be possible to obtain approximate stability parameters in a stable manner using various formulae presented in ref \citenum{McConnell2017microcanonical}.
These approximations, however, lead to systematic errors in the rate \red{constant} and here we are interested in finding a method which does not make any further approximations to instanton theory.
A formally exact, but more stable approach, for calculating the instanton partition function shall therefore be investigated.

\subsubsection{Pinned Ring Polymer} \label{PRP}

A third approach for calculating the instanton vibrational partition function 
\red{was derived by taking a steepest-descent integral over all ring-polymer beads except one. A second steepest-descent integral was then performed over the remaining bead, which we call a ``pinning point'',}
such that the expression depends on derivatives of the action with respect to this pinning point \cite{Althorpe2011ImF,InstReview}
\red{\begin{equation}
\begin{split}
Z^{\mathrm{vib}}_\mathrm{RPI}& = \beta \sqrt{2 \pi \hbar} \left\lVert \dot{x} \right\rVert \left( \frac{-C_{\mathrm{vib}}}{\det' \pders{S}{x}{x}} \right) ^{1/2}
\label{eqn:Z_trajC}
\end{split}
\end{equation}}

A distinction from the RPE and SP methods is that this method uses the topology of a pinned ring polymer in which the first and last beads are not connected by ring-polymer springs.
This means there are no zero eigenvalues corresponding to cyclic permutation of the beads.
\orange{For numerical stability, we will choose} the pinning point, denoted by both $x'$ and $x''$, such that it is located at the bead with the maximum potential energy along the instanton.

Here $ \left\lVert \dot{x} \right\rVert $ is the speed of the trajectory at the pinning point given by the mass-weighted norm of the momentum $\frac{1}{2m} \left\lVert \pder{S}{x''} - \pder{S}{x'} \right\rVert$ and 
\red{$\pders{S}{x}{x} = \pders{S}{x'}{x'} + \pders{S}{x'}{x''} + \pders{S}{x''}{x'} + \pders{S}{x''}{x''}$}. The prime in eq \ref{eqn:Z_trajC} indicates that the determinant in the denominator omits the zero eigenvalues corresponding to translations, rotations, and motion along the reaction coordinate.
The $f_0$ translations and rotations are also dealt with in the numerator according to $C_{\mathrm{vib}} = C \left(\frac{\tau}{m} \right) ^{f_0}$ since the eigenvalues of $- \pders{S}{x'}{x''}$ for a free particle are $m / \tau$.

Using this pinned ring polymer and calculating $C$ using eq \ref{eqn:C_mat}
can be used to obtain thermal reaction rates \orange{except for very low temperatures where this approach becomes numerically unstable}. 
The algorithm described in \Ref{GoldenRPI} 
involves solving a set of linear equations using the banded form of the pinned ring-polymer Hessian, $\mathbf{J}$. \cite{GoldenRPI}
As there are no off-diagonal terms connecting the first and last beads, it is an \orange{$(N-1)f \times (N-1)f$} matrix with bandwidth $f$.
However at low temperatures, $C$ exhibits the same numerical problems seen in the SP methods.
This is because here we also deal with a non-symmetric matrix, $\pders{S}{x'}{x''}$, whose eigenvalues vary by many orders of magnitude.

Another approach for obtaining $C$ is to use the Gelfand-Yaglom (GY) method \cite{Kleinert}.
This uses a recursive approach to obtain
a non-symmetric matrix whose determinant is equal to \red{$\det \mathbf{J}$}.
It is then possible to obtain $C$ from $\mathbf{J}$ using \cite{Althorpe2011ImF}
\begin{equation}
C = \left(\frac{m}{\tau_N}\right)^f (\det\mathbf{J})^{-1}
\label{eqn:Pinned RPE}
\end{equation}
where $\tau_N = \tau/N$.
\red{We find that this method for calculating $C$} has the same problems as eq \ref{eqn:C_mat}.
The failure of these approaches is again attributed to the fact that they attempt to diagonalize a non-symmetric matrix, whose eigenvalues are different from each other by many orders of magnitude at low temperatures.
The matrix \red{$\pders{S}{x}{x}$} does not exhibit this problem because \orange{its spread of eigenvalues is small in magnitude and it is a symmetric matrix, making the calculation of these eigenvalues numerically stable.}

\red{Alternatively, }
we could calculate the determinant of $\mathbf{J}$ as a product of its eigenvalues,
which are found by a banded-matrix eigenvalue solver.
We call this approach ``pinned ring-polymer eigenvalues'' (Pinned RPE)
and
as $\mathbf{J}$ is a symmetric matrix,
this approach is expected to be numerically stable at all temperatures.
However, it scales as approximately $\mathcal{O}(N^2f^3)$ \cite{IntelMKL, KaufmanScaling, LAPACK},
and may still be the bottleneck for an instanton \orange{partition function} calculation.

\begin{figure}
\includegraphics[keepaspectratio, width=3.0in]{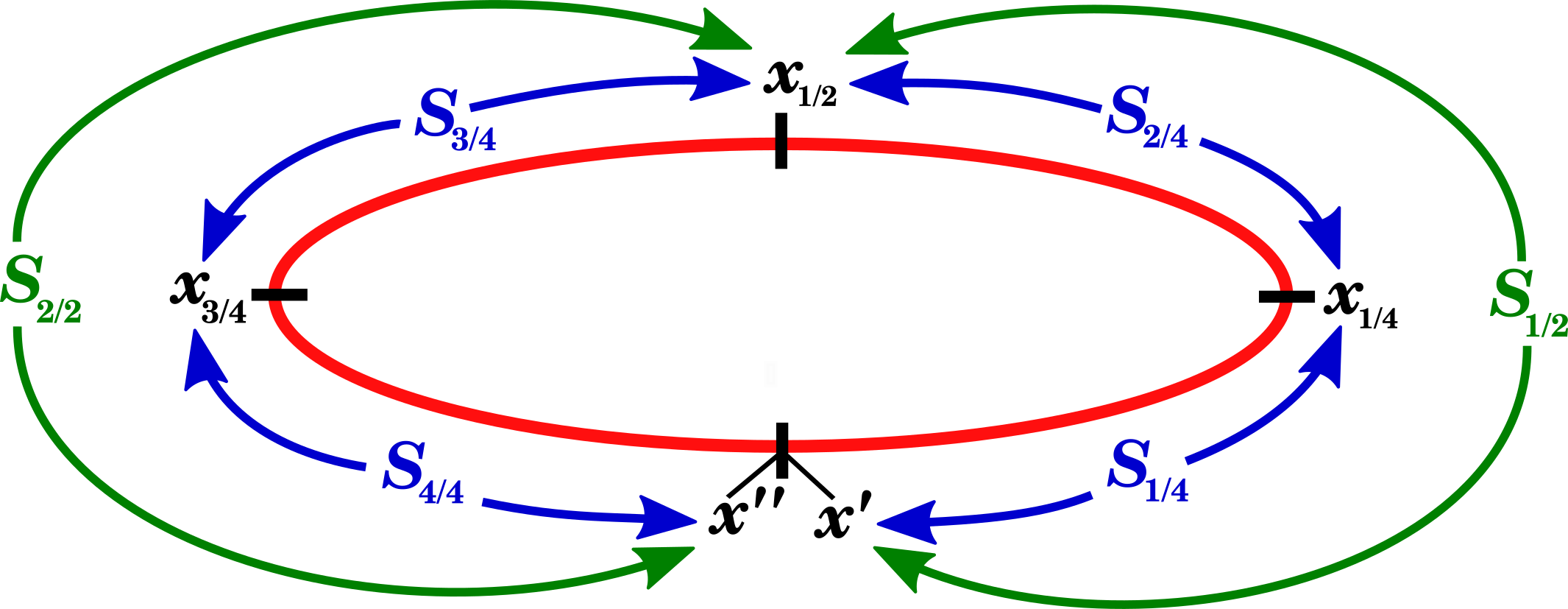}
\caption{Depiction of a full instanton trajectory (red) and its shorter trajectories
after splitting into 2 (green) or 4 (blue) pieces of equal imaginary time $\tau$.}
\label{fgr:DaC_split}
\end{figure}

A method for solving both the problems of low-temperature instability and high computational overhead can be obtained by reconsidering how the trajectories themselves are treated.
If we revisit the semiclassical approximation to the kernel, we see that a trajectory can be split piecewise into two parts giving $ K(x', x'', \tau) = \int K(x', x_0, \tau_a) K(x_0, x'', \tau_b) \, \mathrm{d}x_0$
in which the imaginary time of each piece is additive to give $\tau = \tau_a + \tau_b$ and $x_0$ is the point at which the full trajectory is split.
The kernels for the new shorter trajectories are defined in terms of the actions $S_a = S(x', x_0, \tau_a)$ and $S_b = S(x_0, x'', \tau_b)$.
The choice of $\tau_a$ is arbitrary
and the integral is performed by the method of steepest-descent
around the point $x_0$, which is chosen such that the particle's momentum is continuous to give $\pder{S_a}{x_0} + \pder{S_b}{x_0} = 0$.
For simplicity however, we will choose values of $\tau_a=\tau/2$ which split the initial trajectory exactly in half,
and define $x_0 = x_{1/2}$ to obey the condition.
In order to denote this splitting, the subscripts $a$ and $b$ of the action will become $1/2$ and $2/2$ respectively. We can then redefine $C$ according to \cite{GutzwillerBook}
\begin{equation}
C = C_{1/2} \left|  \frac{\partial^2{S_{1/2}}}{\partial{x_{1/2}^2}} +  \frac{\partial^2{S_{2/2}}}{\partial{x_{1/2}^2}}\right|^{-1} C_{2/2} 
\label{eqn:C_split2}
\end{equation}
whereby the determinant $C$ for the full instanton trajectory is obtained using the determinants of the smaller halves and the cross-term between them.
The determinants of the shorter trajectories are defined as follows:
\begin{equation}
C_{1/2} = \left| - \pders{S_{1/2}}{x'}{x_{1/2}} \right| \qquad \text{and} \qquad C_{2/2} = \left| - \pders{S_{2/2}}{x_{1/2}}{x''}\right|
\label{eqn:C_halves}
\end{equation}
A particularly useful characteristic of this new ``Divide-and-Conquer'' (DaC) approach is that it can be applied multiple times in order to further stabilize computations with large $\tau$. For example, splitting a trajectory into 4 pieces can be performed by simply reapplying eq \ref{eqn:C_split2} to both $C_{1/2}$ and $C_{2/2}$ giving
\begin{equation}
C =  \underbrace{ C_{1/4} \left|  \frac{\partial^2{S_{1/4}}}{\partial{x_{1/4}^2}} +  \frac{\partial^2{S_{2/4}}}{\partial{x_{1/4}^2}}\right|^{-1}  C_{2/4} }_{\mathlarger{C_{1/2}}}
\left|  \frac{\partial^2{S_{1/2}}}{\partial{x_{1/2}^2}} +  \frac{\partial^2{S_{2/2}}}{\partial{x_{1/2}^2}}\right|^{-1}
 \underbrace{ C_{3/4} \left|  \frac{\partial^2{S_{3/4}}}{\partial{x_{3/4}^2}} +  \frac{\partial^2{S_{4/4}}}{\partial{x_{3/4}^2}}\right|^{-1}  C_{4/4}}_{\mathlarger{C_{2/2}}}
\label{eqn:C_split4}
 \end{equation}
This splitting is represented schematically in Figure \ref{fgr:DaC_split}, in which the red path corresponds to the full trajectory of the instanton which begins at $x'$ and ends at $x''$. 
The approach which calculates $C$ using the full instanton trajectory, as in eq \ref{eqn:C_mat}, will be called DaC1.
The green terms in Figure \ref{fgr:DaC_split} show the splitting of the full trajectory into 2 pieces, as in eq \ref{eqn:C_split2}, and this approach will be called DaC2. The blue terms in Figure \ref{fgr:DaC_split} show the splitting of the full trajectory into 4 pieces, as in eq \ref{eqn:C_split4}, and this approach will be called DaC4.
Note that the DaC4 method requires information about the DaC2 trajectories and hence also about the full DaC1 trajectory.
Once the number of splits is large enough,
all terms in the expression \red{for C} become numerically stable.
\red{Compared to the full trajectory, the split DaC trajectories have shorter path lengths and travel in a fraction of the imaginary time.
For this reason, changes in the initial positions and momenta of the trajectories with shorter path lengths will have less of an impact on the corresponding final positions and momenta.
This means that derivatives of the action with respect to the starting and ending bead positions are smaller in magnitude, leading to a smaller spread of eigenvalues and therefore more numerical stability during the calculation of $C$.}
The ring-polymer vibrational partition function, and hence the reaction rate,
can then be obtained \orange{in a numerically stable manner at very low temperatures.}

\orange{Note that} this approach is not a generalizable determinant finder for any matrix, but rather a specialized method for calculating the determinant, $C$, of the fluctuations around a trajectory.
The algorithm requires as input all bead positions, gradients, and Hessians in order to calculate the necessary derivatives of the action \cite{GoldenRPI}.

\section{Results and Discussion} \label{Results and Discussion}

Comparisons between all the approaches described in section \ref{Zvib}
were performed by calculating thermal \red{rate constants} for three different potential-energy surfaces.
The results show in each case that the DaC$n$ method is efficient and stable even at low temperatures,
as long as $n$ is chosen large enough.

\subsection{\ch{H + CH4} on a Fitted PES}

\begin{figure}[H]
\includegraphics[keepaspectratio, width=3.33in]{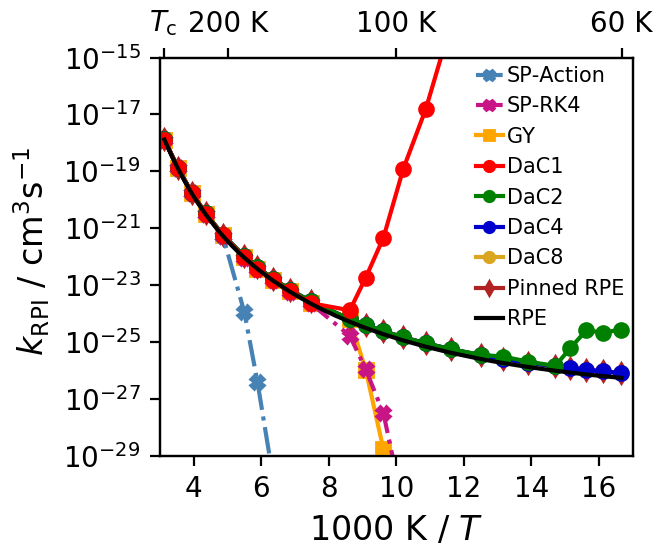}
\caption{Instanton reaction rates, $k_{\mathrm{RPI}}$, for the bimolecular \ch{H + CH4} reaction using the LCZXZG PES with crossover temperature $T_\mathrm{c} = 331\text{ K}$. Instantons were optimized using $N=512$.
Markers for the Pinned RPE and DaC8 methods are obscured by those of DaC4, which have almost exactly equal values.
}
\label{fgr:HCH4_rates}
\end{figure}

The \ch{H + CH4} reaction is a useful test system for determining bimolecular reaction rates with chemical accuracy.
\cite{Wu2004HCH4,Welsch2012HCH4,Suleimanov2011HCH4,Andersson2009Hmethane,HCH4,MUSTreview}
The LCZXZG potential-energy surface was fitted using a polynomial neural network 
to describe the full 18-dimensional 
reaction \cite{Li2015LCZXZG}.
It includes translational and rotational degrees of freedom and has a crossover temperature of $T_\mathrm{c} = 331~\mathrm{K}$\@.
Gradients and Hessians of the PES were computed using finite differences.
Ring-polymer instantons were optimized for this surface using 512 beads and a convergence criterion such that the norm of the gradient of the ring-polymer potential was less than $10^{-5} \ \text{eV/\AA}$.
Thermal \red{rate constants} were then calculated for the different approaches outlined in Figure \ref{fgr:HCH4_rates}.
Because this is a bimolecular system, which does not have a low-temperature plateau,
$\frac{\mathrm{d}E}{\mathrm{d}\tau}$ has been calculated as the \red{total} derivative, eq \ref{eqn:d2WdE2}, and has not been approximated.
We find that this part of the calculation is numerically stable
and gives equivalent results to the simple finite-difference approach.
To take account of the reaction symmetry, we use $\sigma=4$.

The SP-Action method is seen to break down near 200 K and this is attributed to numerical errors
emanating both from building the monodromy matrix, $\mathbf{M}(\tau)$, and from diagonalizing it. The SP-RK4, DaC1, and GY methods all diverge at approximately 120 K which suggests that here the failure is due to the diagonalization of the respective non-symmetric matrices.
At least for this system,
the Runge-Kutta integration scheme appears to be more stable in building $\mathbf{M}(\tau)$ than the SP-Action approach.
Nevertheless, all methods based on stability parameters exhibit numerical problems which can not be fixed by simply increasing the number of beads. 

As the temperature decreases and the instanton stretches over a larger trajectory, the DaC splitting methods are able to avoid these numerical problems by treating the trajectory in a piecewise manner. Figure \ref{fgr:HCH4_rates} shows how increasing the number of splits within the DaC method allows for convergence of the rate \red{constant} even at very low temperatures. The values from both the DaC8 and Pinned RPE methods are hidden behind the DaC4 points for all temperatures, suggesting that these methods give the same numerical values of $C$. 
At the lowest temperatures studied there exists a minor ($\sim$10\%) deviation between the DaC$n$ (for $n\ge4$) and RPE methods. The origin of this difference is not clear, as both methods are in principle equivalent, and is likely due to a small remaining numerical error in either RPE or DaC$n$.
This error is however lower than the expected error introduced by the semiclassical approximation used in RPI theory and is therefore not of large concern.
More importantly, the DaC method is stable across all temperatures as long as the instanton trajectory is split into a sufficient number of pieces.

In summary, three methods are able to calculate accurate \ch{H + CH4} reaction rates from the crossover temperature $T_\mathrm{c}$ down to at least 60K: RPE, Pinned RPE, and the newly proposed DaC method.
We have no reason to believe that this behavior will not continue to even lower temperatures.

\begin{figure}[H]
\includegraphics{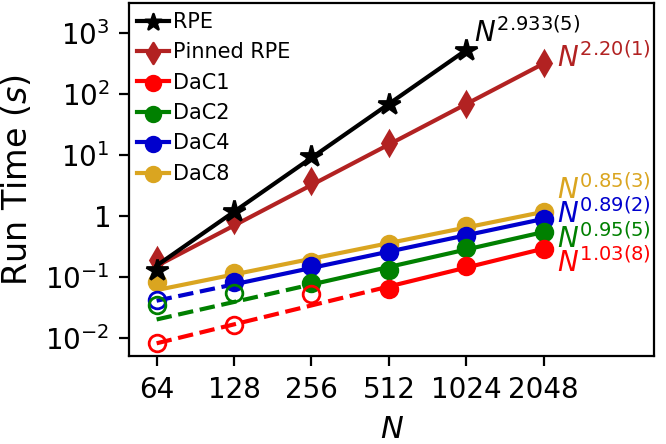}
\caption{Computational run time and scaling required to calculate $Z^{\mathrm{vib}}_\mathrm{RPI}$ as a function of $N$ for the \ch{H + CH4} reaction. Lines of best fit were modeled using power functions and optimized using non-linear least squares.
Points with open circles were omitted from the lines of best fit due to large variability at such a short time scale.
\red{The numbers in parentheses represent a standard deviation in the last digit.}
Run times were measured on a standard desktop computer with a single processor.
}
\label{fgr:HCH4_scaling}
\end{figure}

In order to compare the computational requirements for these methods, run times for evaluating $Z^{\mathrm{vib}}_\mathrm{RPI}$ were measured using the same computer without the use of parallelization nor sparse matrix \red{algorithms}.
\red{Note that these run times do not include the time required to locate the instanton nor to compute the Hessians at each bead.}
Figure \ref{fgr:HCH4_scaling} compares the scaling of these three accurate methods with respect to the number of beads, $N$. 
It is clear that the run times for the RPE method are likely to be too large to be tractable for systems requiring many beads.
For 2048 beads, the memory demand is already too large to store the $Nf \times Nf$ matrix on our desktop computer, although \red{employing sparse matrix algorithms} could alleviate this problem.
As discussed in section \ref{RPE}, diagonalizing a matrix of this size will scale approximately as $\mathcal{O}(N^3f^3)$, \cite{IntelMKL, KaufmanScaling, LAPACK} which can become too expensive to calculate reaction rates for systems of interest.

Since the Pinned RPE method describes a pinned rather than cyclic ring polymer, the $\mathbf{J}$ matrix is banded with bandwidth $f$.
Generalized banded matrices can be diagonalized more efficiently with a theoretical scaling of approximately $\mathcal{O}(N^{2}f^3)$ \cite{IntelMKL, KaufmanScaling, LAPACK}, which is in agreement with our study. 

The proposed DaC method avoids
performing costly matrix diagonalizations,
and instead its computational bottleneck is in solving $f$ sets of $Nf$ linear equations defined by the banded matrix $\mathbf{J}$.
This has a scaling of approximately $\mathcal{O}(Nf^3)$ \cite{NumRep}, enabling the DaC method to be used for larger systems with many beads.
Importantly, splitting the trajectory multiple times does not affect the scaling for calculating $Z^{\mathrm{vib}}_\mathrm{RPI}$ with respect to the number of beads.
For calculations with 2048 beads, the RPE method requires a run time on the order of hours, the Pinned RPE method requires a run time on the order of minutes, and our new DaC methods require run times of less than one second. 
Our method can thus lead to a significant speed up for instanton rate calculations at low temperatures where many beads are needed.
This reduction in run time and scaling will be particularly pronounced if using cheap potential-energy surfaces, such as those built by machine-learning approaches.

\subsection{Analytic M{\"u}ller-Brown Potential} \label{ssec:MB}

The M{\"u}ller-Brown surface is a 2-D model commonly used to test new theoretical approaches \cite{Muller1979}. We employ it here to demonstrate that the problems we highlight are endemic and exist even for this simplest system,
where other numerical errors can be eliminated.
For example, a very large number of beads could be used and the convergence criteria for instanton optimization was set such that the norm of the gradient was less than $10^{-8}  \ E_\text{h}/a_0$.
The calculations of gradients and Hessians were performed analytically and this system has neither translational nor rotational modes which require special treatment.
Here we investigate the reaction from the local minimum with coordinates $(-0.05001, 0.46669)$ through the saddle point at $(-0.822001, 0.624314)$ to stay consistent with calculations run by \citeauthor{KastnerAction} \cite{McConnell2017instanton, KastnerAction}.
Note that the PES is also scaled in the same way such that the barrier height between these two points is $0.19\ E_\mathrm{h}$,
which gives a crossover temperature of $T_\text{c} = 2210 \text{ K}$.
Instantons were located for this reaction in order to test how the different methods for calculating $Z^{\mathrm{vib}}_\mathrm{RPI}$ converge as a function of the number of beads, $N$.
Since there are neither translations nor rotations, the corresponding partition functions in eqn \ref{eqn:kt} are simply equal to 1.

\begin{figure}
\includegraphics{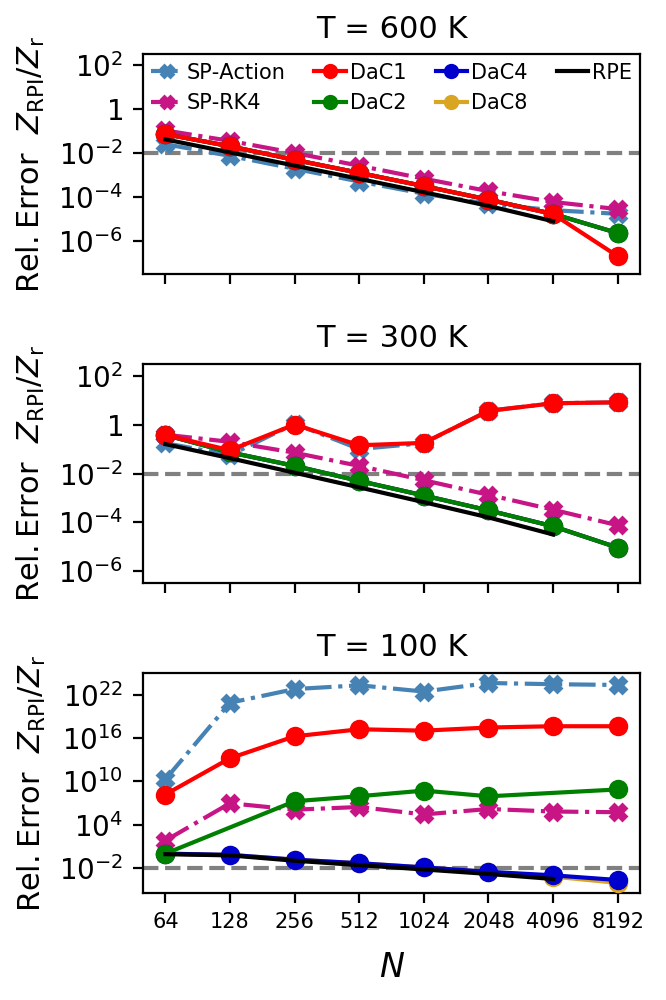}
\caption{\red{Relative error in calculating $Z_{\mathrm{RPI}} / Z_{\mathrm{r}}$ as a function of the number of beads, $N$, at three different temperatures.} 
\red{The relative error was defined using the RPE results with $N=8192$ as the benchmark for each respective temperature as $\left| \frac{X_{\mathrm{method}, N}-X_{\mathrm{RPE}, 8192}}{X_{\mathrm{RPE}, 8192}} \right |$,
where $X$ denotes $Z_\mathrm{RPI} / Z_\mathrm{r}$ for a given method and number of beads, $N$.
No points are plotted for the benchmark because it has zero relative error by definition.}
\red{A horizontal dashed line indicates a 1\% relative error, below which results are considered converged.}
In all cases, DaC8 results are so similar to DaC4 results that they cannot be distinguished on the plot.
}
\label{fgr:MB}
\end{figure}

In Figure \ref{fgr:MB} we plot the \red{relative error} in calculating the ratio of total partition functions $Z_\mathrm{RPI} / Z_\mathrm{r}$ with respect to our benchmark, the RPE method with $N=8192$.
The Pinned RPE method gives values which are identical to those of the DaC8 method and are therefore not plotted here. Note that in all parts of Figure \ref{fgr:MB}, the precedence of points is plotted such that the DaC$n$ methods with small $n$ are plotted on top of those with successively higher values of $n$.
The SP methods were computed using the approximation $\frac{\mathrm{d}E}{\mathrm{d}\tau}  \approx \pder[2]{S}{\tau}$, which is valid here since we are in the plateau region where the unimolecular rate is independent of temperature.

All methods converge to within much less than a 1\% error of the benchmark values at 600~K\@.
When the temperature is decreased to 300 K however, the SP-Action and DaC1 methods break down and fail to converge for any number of beads, $N$.
The DaC$n$ (for $n \ge 2$) and SP-RK4 methods converge smoothly to the benchmark result at approximately the same rate as the RPE method.

At 100 K, the SP-Action and SP-RK4 as well as the DaC1 and DaC2 methods all break down completely, regardless of the number of beads used. 
In some cases, e.g. the DaC2 method with $N=128$ and $N=4096$, $C$ even has the wrong sign such that the instanton vibrational partition function cannot be computed.
By splitting the trajectory into smaller pieces, however, these stability and convergence problems can be remedied.
The DaC4 and DaC8 methods which split the trajectory into many pieces converge smoothly with respect to $N$
to the benchmark result just as quickly as the RPE method.
The DaC$n$ methods are also seen to converge with respect to $n$ as using a higher number of splittings does not affect the results.
This is shown by the fact that DaC8 values are equal to DaC4 values for all number of beads and across all three temperatures.

Note that in order to obtain the results presented above,
it was necessary 
to choose the pinning locations $x'$ and $x''$ to be at the bead of maximum potential energy.
It was particularly important in this low-temperature regime
as most of the beads are located in the reactant well.
The speed, $\left\lVert \dot{x} \right\rVert$, of these beads is approximately zero,
which would not lead to a numerically stable method for calculating eq \ref{eqn:Z_trajC}.
However, using the bead with maximum potential energy was found to be reliable throughout.

Although formally equivalent, the methods presented here show large numerical deviations from each other. These results clearly show that the stability parameters methods and DaC$n$ methods with small $n$ cannot be used to calculate the instanton vibrational partition function at low temperatures in the deep tunneling regime.
Under these conditions, increasing the number of beads does not help converge $Z_\mathrm{RPI}$ with respect to results obtained by the benchmark method.
Note that \citeauthor{KastnerAction} concluded that the SP methods were stable for this system
because they did not consider such low temperatures \cite{KastnerAction}.
Our study shows that the SP methods break down due to numerical instability in the diagonalization of non-symmetric matrices, as discussed in section \ref{stabil}.

\subsection{On-the-fly Criegee Intermediate}
Criegee intermediates play an important role in producing \ch{OH} radicals throughout Earth's atmosphere \cite{Chhantyal-Pun2017, Novelli2014, Stone2012, Alam2011, Donahue2011}.
The unimolecular hydrogen transfer reaction of the syn-\ch{CH3CHOO} Criegee intermediate can determine reaction rates and branching ratios relevant for \ch{OH} production. Moreover, this reaction is known to proceed via a tunneling mechanism
\cite{Fang2016CH3CHOO}
and can therefore be well characterized using RPI theory.

\begin{figure}
\includegraphics[keepaspectratio, width=2.0in]{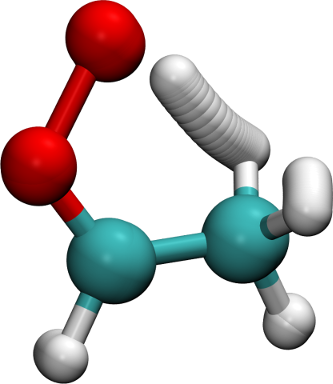}
\caption{Depiction of an instanton for the hydrogen transfer reaction of the syn-\ch{CH3CHOO} Criegee intermediate with $T_\mathrm{c} = 477 \text{ K}$. This instanton was optimized at 100 K with $N=64$.}
\label{fgr:Criegee}
\end{figure}

We use an on-the-fly \orange{ab initio} PES, locate instantons, and compute reaction rates from 400 K to 100 K for this hydrogen transfer reaction.
DFT electronic-structure calculations were performed with Molpro \cite{MOLPRO-WIREs} using the B3LYP hybrid functional and the cc-pVDZ basis set \cite{Becke1988, Lee1988, Vosko1980}.
Single-point energy calculations were performed using the default Molpro convergence threshold of $10^{-6} \ E_\mathrm{h}$.
Gradients of the potential energy are calculated analytically and Hessians are calculated by finite differences.
\red{In this system, the bottleneck for computation time was the calculation of the Hessians at each bead rather than the vibrational partition function,
and the crossover temperature was found to be 477 K\@.}
Instantons were optimized on this surface using 64 beads and a gradient convergence threshold of $10^{-5}  \ E_\mathrm{h}/\orange{\text{\AA}}$.
The instanton optimized at 100 K is shown in Figure \ref{fgr:Criegee} and it should be noted that at this temperature,
the system is in the plateau region. 

The results for the different rate \red{constant} methods using these instantons are presented in Table \ref{tbl:Criegee_OTF}, \orange{where the symmetry factor for the reaction, $\sigma = 2$, cancels out in the numerator and denominator.}
The thermal rates calculated using RPI theory with $N=64$ are presented relative to the standard Eyring TST thermal rates in which no finite-bead approximation was used.
The SP methods here were computed using $\frac{\mathrm{d}E}{\mathrm{d}\tau}$ as defined in eq \ref{eqn:d2WdE2}.
As well as the problems for obtaining the stability parameters,
the calculation of $\frac{\mathrm{d}E}{\mathrm{d}\tau}$ is not reliable and a slight change in the bead index $i$ of eq \ref{eqn:d2Sdt2} leads to significant changes in magnitude and even the sign of $\frac{\mathrm{d}E}{\mathrm{d}\tau}$ \orange{at very low temperatures.
Using the approximation $\frac{\mathrm{d}E}{\mathrm{d}\tau}  \approx \pder[2]{S}{\tau}$ does not help in this case,
which is yet another
reason to avoid calculating instanton partition functions using the SP methods.}

\begin{table}[H]
  \caption{Quantum tunneling factors, defined by the ratio of instanton to Eyring TST rates, $k_\text{RPI} / k_\text{Eyring}$,
  for the syn-\ch{CH3CHOO} hydrogen transfer.
  Instantons were optimized on-the-fly at the B3LYP/cc-pVDZ level with $N=64$. \orange{Numbers in parentheses denote powers of 10.}}
  \label{tbl:Criegee_OTF}
  \begin{tabular}{c c c c}
    \hline
    Method & 400K & 200K  & 100K \\
    \hline
    SP-Action & 45.52 & 1.29(12) &  6.72(14)\\
    SP-RK4 & 44.10 & 1.28(12) & 2.14(37) \\
    DaC1 & 45.20 & 1.57(12) & 1.20(44) \\
    DaC2 & 45.12 & 1.54(12) & 1.36(41) \\
    DaC4 & 45.12 & 1.54(12) & 1.36(41) \\
    DaC8 & 45.12 & 1.54(12) & 1.36(41) \\
    Pinned RPE & 45.12 & 1.54(12) & 1.36(41) \\
    RPE & 45.23 & 1.54(12) & 1.42(41) \\
    \hline
  \end{tabular}
\end{table}

Due to many numerical approximations, introduced for example by the iterative scheme inherent in the electronic-structure calculation, an on-the-fly \orange{ab initio} surface can be much less smooth than a fitted or analytic PES\@.
We see in Table \ref{tbl:Criegee_OTF} that the DaC$n$ methods appear to be stable despite the ``roughness'' of this \orange{ab initio} surface. 
As the temperature decreases, the SP and DaC1 methods again break down when compared to the benchmark RPE method. 
The DaC$n$ (for $n\ge2$) methods give the same values as the Pinned RPE method at all temperatures and, as in the \ch{H + CH4} system, there is a minor ($\sim$5\%) discrepancy between these methods and the benchmark.
This is again due to unidentified numerical errors, which do not overly concern us as they are so small,
and it is not clear whether the benchmark itself is at fault.

In this case, the error introduced by the electronic-structure calculation is most likely larger than the instanton error.
The goal of this calculation is therefore a proof of concept to show that the DaC method within instanton theory can be used to calculate thermal reaction rates on the fly.
Accuracy of the \red{rate constants} could be improved by using a higher-level electronic-structure theory for constructing the surface, for instance by using a machine-learning PES trained on coupled-cluster calculations
\cite{Bowman2015CH3CHOO,Kidwell2016CH3CHOO}. Incorporating a more accurate PES would not change the conclusions drawn here, however. The DaC method would continue to be stable and the stability parameters methods would still exhibit numerical problems at low temperature.

\section{Conclusion}

Ring-polymer instanton reaction rates have been calculated for three different potential-energy surfaces: a fitted 18-D PES for a bimolecular reaction, a 2-D analytic model, and a 24-D unimolecular reaction calculated on the fly.
All the methods for obtaining the instanton partition function discussed in this paper are formally equivalent
but differ in how they are calculated numerically.

Many of the current methods for obtaining the instanton vibrational partition function were tested on these surfaces and they fall under two classes. The first class includes the RPE and Pinned RPE methods, which are able to calculate accurate rates at low temperatures and they exhibit numerical stability and convergence with respect to the number of beads, $N$.
\red{However, poor scaling and long computation time make these approaches unfavorable when calculating rate constants \orange{at very low temperature} and when using a large number of ring-polymer beads.}
The second class of methods used to calculate the instanton vibrational partition function include the SP, GY, and DaC1 methods. These approaches are faster and scale better than the methods in the first class but can \orange{not} be used reliably at \orange{very low temperatures.
In this regime}, they exhibit many numerical problems regardless of the number of beads used to discretize the instanton trajectory.

Our proposed DaC$n$ method (once converged with respect to $n$) is an advancement for thermal instanton theory because it is able to calculate reaction rates at all temperatures \orange{below crossover} accurately and with favorable scaling. 
It has shown numerical stability across three types of potential-energy surfaces and can be used efficiently for systems at \orange{very} low temperature with a large number of ring-polymer beads, qualities which are not true for any of the previous methods.

The Divide-and-Conquer method will be particularly useful when coupled with methods which obtain the PES cheaply, such as machine-learning approaches, because in these cases the bottleneck of the whole calculation may be the time required to calculate the instanton vibrational partition function. 
Even for realistic examples where an \orange{ab initio} PES is calculated on the fly, this method has shown to be stable and reliable for calculating thermal reaction rates. Our method significantly reduces the challenge in calculating thermal reaction rates of larger chemical systems at \orange{very} low temperatures \red{where a large number of ring-polymer beads are required}.
\cite{porphycene}

\begin{acknowledgement}

This work was financially supported by the Swiss National Science Foundation through SNSF Project 175696.

\end{acknowledgement}

%
%
%
%
\bibliography{../bibliographies/references,../bibliographies/pierre_references}

\newpage
\begin{figure}
\includegraphics[keepaspectratio, width=2.0in]{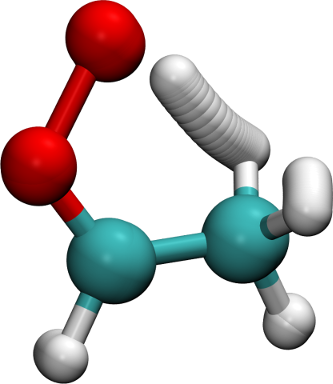}
\caption*{For Table of Contents Only}
\label{For Table of Contents Only}
\end{figure}

\end{document}